\documentclass{svjour3}

\usepackage{amssymb,amsmath,amsfonts}
\usepackage{color}
\usepackage{epsfig}
\usepackage{epstopdf}
\usepackage{fancyhdr}
\usepackage{float}
\usepackage{graphicx}
\usepackage{times}
\usepackage{url}

\def\thercsid{\relax}

\renewcommand{\today}{\number\day\space\ifcase\month\or
  January\or February\or March\or April\or May\or June\or
  July\or August\or September\or October\or November\or December\fi
  \space\number\year}

\def\be{\begin{equation}}
\def\ee{\end{equation}}
\def\bi{\begin{itemize}}
\def\ei{\end{itemize}}
\def\ben{\begin{enumerate}}
\def\een{\end{enumerate}}

\def\Msun{\mathrm{M}_\odot}

\begin{document}

\title{Multimessenger astronomy with the Einstein Telescope%\thanks{Grants or other notes
}


% http://www.aip.org/pacs/pacs08/pacs0800.html
%%\pacs{
%%04.80.Nn, % Gravitational wave detectors and experiments
%%04.80.Cc  % Experimental tests of gravitational theories
%%}

%\date[\relax]{RCS \thercsid; compiled \today }

% Authorlist alphabetical ...................................................

\newcommand*{\AUTHORone}{Eric~Chassande-Mottin}
\newcommand*{\UNIone}{AstroParticule et Cosmologie (APC), CNRS – IN2P3 – Observatoire de Paris – Universit\'e Denis Diderot - Paris 7 – CEA, 75013 Paris, FRANCE\\ Tel.: 33 15727 6036\\ Fax: 33 15727 6071\\\email{ecm@apc.univ-paris7.fr}}
\newcommand*{\AUTHORtwo}{Martin~Hendry}
\newcommand*{\UNItwo}{Dept of Physics and Astronomy, University of Glasgow, Glasgow, G12 8QQ, UK\\ Tel.: 44 141 330 5685\\ Fax: 44 141 330 5183\\\email{martin@astro.gla.ac.uk}}
\newcommand*{\AUTHORthree}{Patrick~J.~Sutton}
\newcommand*{\UNIthree}{School of Physics and Astronomy,
Cardiff University, Cardiff, UK, CF24 3AA\\ Tel.: 44 292 087 4649\\ Fax:  44 292 087 4056\\\email{patrick.sutton@astro.cf.ac.uk}}
\newcommand*{\AUTHORfour}{Szabolcs~M\'{a}rka}
\newcommand*{\UNIfour}{Columbia Astrophysics Laboratory,
Columbia University in the City of New York, Pupin Physics Laboratories, New York, NY 10027 (USA)\\ Tel.: 1 212 854 8209\\ Fax: 1 212 854 8121\\\email{sm2375@columbia.edu}}

\author{\AUTHORone \and \AUTHORtwo \and \AUTHORthree \and \AUTHORfour}

\institute{\AUTHORone \at \UNIone \and
  \AUTHORtwo   \at \UNItwo \and
  \AUTHORthree \at \UNIthree \and
  \AUTHORfour  \at \UNIfour
}

% Don't forget about others...
% .............................................

\maketitle

\begin{abstract}
\vspace*{0.2in}
Gravitational waves (GWs) are expected to play a crucial r\^ole in the
development of multimessenger astrophysics. The combination of GW observations
with other astrophysical triggers, such as from gamma-ray and X-ray satellites,
optical/radio telescopes, and neutrino detectors allows us to decipher science that
would otherwise be inaccessible. In this paper, we provide a broad review from
the multimessenger perspective of the science reach offered by the third generation
interferometric GW detectors and by the Einstein Telescope (ET) in particular. 
We focus on cosmic transients, and base our estimates on the
results obtained by ET's predecessors GEO, LIGO, and Virgo.

\end{abstract}

%\begin{keyword}
%multi-messenger \sep gravitational wave \sep neutrinos \sep transients
%\end{keyword}

\date[\relax]{ RCS \thercsid; compiled \today }

% .............................................

\maketitle

\section{Introduction} \label{section:Introduction}

Coalescing binaries, core-collapse supernovae, and magnetars are not only interesting candidates for
gravitational wave (GW) searches, but are also observed by other means, such
as gamma-rays, X-rays, visible light, radio waves, and neutrinos. Therefore GW
science in particular and astrophysics in general can profit from joint
observations of astrophysical events detected by multiple observatories.
Even simple correlation in time and direction between different messengers
that correspond to the same astrophysical event can greatly increase the
confidence in a detection of GWs, and search strategies can be optimized
in this respect. Furthermore, several long-term goals of GW astrophysics
require detection of astrophysical events in other channels beyond GWs.
For example, an association between short hard GRBs and inspiralling
neutron star binaries may be confirmed in this manner. The joint detection
of GWs and neutrinos together with the observation of the optical light
curve from a nearby supernova would greatly enhance our understanding of
supernova explosions. Thanks to their ten-fold improvement in sensitivity,
the third generation of interferometric GW detectors and the Einstein Telescope
(ET) in particular will allow valuable astrophysical statements to be made through
multimessenger observations. In this paper, we present an overview of
the science reach that can be attained this way, focusing on astrophysical 
transients\footnote{A detailed overview of continuous GW signals is given in 
reference\,\cite{2009astro2010S.229O,Owen:2009tj} and in the article by
Andersson {\em et al.\/} in this volume.}.
Our goal is not to be complete in our coverage, but instead to highlight 
a sample of the science topics that will benefit from a multimessenger 
approach -- emphasising both the wide range and the potential impact of 
these topics.  

Before beginning an examination of particular astrophysical systems,
it is worthwhile to consider how the nascent field of GW astronomy has
interacted with traditional electromagnetic (EM) astronomy to date.
There has not yet been a direct detection of GWs; expected GW signal strengths
are weak compared to the background noise levels of current detectors, and searches
are hampered by the non-stationary (``glitchy'') nature of that background noise.  With this in mind, it is perhaps not surprising that the application of other messengers (mainly EM observations) in GW astronomy has been primarily with an eye to making a first detection.  Information obtained from EM observations is used to improve the sensitivity of GW searches,
and to increase the confidence of a putative GW candidate.  For example,
knowing the time of an astrophysical event permits a focussed GW search
on a short period of data for an associated gravitational-wave signal,
reducing the false alarm probability of the GW search.  Knowledge of
the sky position (or sometimes other parameters such as frequency, e.g. from
the measurement of quasi periodic oscillations) allows the rejection of background events inconsistent
with those constraints.  An estimated distance to the source (as may be available
for SGRs and GRBs) allows the selection of particularly promising systems for analysis. In this light, the natural mode for multimessenger cooperation is that EM observation of an astrophysical transient {\em triggers\/} a GW search~\cite{4051k} -- this approach has been adopted in many searches by LIGO, Virgo and other GW detectors, particularly searches triggered by observations from gamma-ray and X-ray satellites, e.g.~\cite{GRB030329,2007PhRvD..76f2003A,SGRpaper,070201,abbott:211102,S2S3S4,Collaboration:2009zd,4051k,2009arXiv0908.3824L,Kalmus08Thesis}.

This EM-triggered mode of collaboration is also natural when one considers the
qualitatively different nature of EM and GW observatories.  GW detector networks
are all-sky monitors, with a typical angular resolution of several degrees. The
data are in the form of digitized times-series collected at rather low rates
typically of $\sim$O($10^4$ samples/s).  The low data rate allows all data to be
archived; since pointing is achieved by aperture synthesis, triggered searches can 
be conducted well after
the data are collected.  EM observatories, by contrast, are generally highly
directional, typically with a field of view (FOV) that is arcminutes in scale.
Current and planned future radio arrays such as LOFAR and the Square Kilometre
Array have a much larger FOV, and point via aperture synthesis, but their large
data rate prevents all data from being archived, so in general decisions about
pointing must still be made at the time of observation.

Nascent efforts already exist to conduct multimessenger searches which go
the other way: EM observations~\cite{2009arXiv0902.1527B,2008CQGra..25r4034K}
being conducted as follow-up to GW triggers. This approach 
recognises the potential value of having multimessenger observations of a
candidate GW detection: these can provide independent confirmation of the
signal and assist with its interpretation. Large FOV optical and infrared telescopes
which are already in existence, or which will see first light in the near future,
provide very exciting prospects for such joint observations.  However, these
opportunities come with significant challenges.  Perhaps foremost among
these is the task of imaging areas of O(10~deg$^2$), and from all of the
objects in that FOV identifying the single EM transient associated with
the GW event. Looking ahead to the advanced LIGO / Virgo era, and further to the ET era, we
expect GW detections to be a regular occurrence. In those circumstances multimessenger
observations will be essential to maximize the scientific benefit from
opening the new GW spectrum.

\begin{figure}
\begin{center}
\includegraphics[width=\columnwidth]{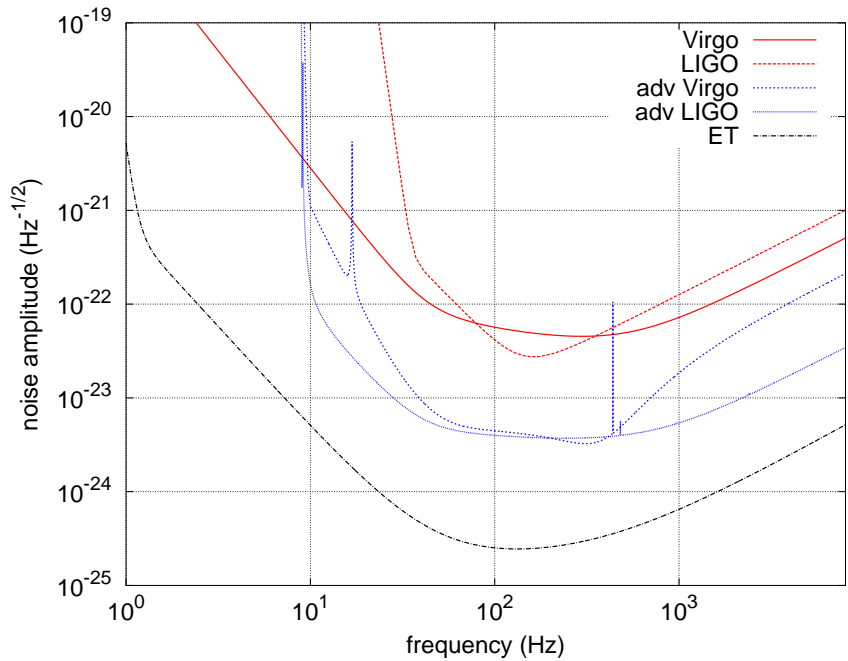}
\caption{Design strain noise amplitude spectra of ET and first- and
second-generation gravitational-wave detectors. The target sensitivity of those instruments are obtained from \cite{punturo04:_virgo,lazzarini96:_ligo_scien_requir_docum,collaboration09:_advan_virgo_basel_desig,shoemaker09:_advan_ligo,hild08:_pushin_et}.}
\label{fig:spectrum}
\end{center}
\end{figure}

Of course we should also mention here neutrino observatories.  These provide an interesting case~\cite{Aso08} for joint GW observations because, as we shall see, the FOV, angular resolution and distance sensitivity are similar to those of GW detectors.

To complete our introductory remarks it is helpful to define a \emph{rule of
  thumb\/} to approximate the distance sensitivity of a generic gravitational
wave detector. Most of the sources considered here do not have well-modelled
gravitational waveforms.  However, if we consider narrowband emission (i.e.,
bandwidths much smaller than the central frequency of the emission, and smaller
than the typical frequency range over which the detector noise spectrum changes), 
we can estimate the distance to which a detector is sensitive as a function of
frequency.  Assuming emission of energy $E_\mathrm{GW}$ in gravitational 
waves\footnote{The details of the assumed polarization content and emission pattern 
affect the result by a factor of O(1) \cite{Sutton:10}, which we can ignore for an 
order-of-magnitude estimate.}
%at frequency $f$, and for simplicity neglecting cosmological effects, 
at source frequency $f_e$, 
the typical sensitive range $D_\mathrm{L}$ (luminosity distance) is approximately 
\cite{Sutton:10}
\begin{eqnarray}\label{eq:DL}
D_\mathrm{L} & \simeq & \sqrt{\frac{G \: (1+z) \: E_\mathrm{GW}}{2 \pi^2 c^3 \: S(f)}} \, \frac{1}{\rho_\mathrm{det} f} \nonumber \\
  & \simeq & 2~\mathrm{Gpc} \, (1+z)^{1/2} \, \frac{10}{\rho_\mathrm{det}} \frac{100 \,\mathrm{Hz}}{f} \left(\frac{E_\mathrm{GW}}{10^{-2}M_\odot c^2} \right)^{1/2} \frac{2.5 \times 10^{-25} \mathrm{Hz}^{-1/2}}{S(f)^{1/2}} \, .  %/\sqrt{\mathrm{Hz}}
\end{eqnarray}
Here $\rho_\mathrm{det}$ is the signal-to-noise ratio in a single detector required for a detection, 
% ---- cosmological effects:
$f=f_e/(1+z)$ is the observed frequency, 
and $S(f)$ is the detector noise power spectrum.  Design noise spectra for
LIGO, Virgo, Advanced LIGO, and ET are shown in Figure~\ref{fig:spectrum}.

The remainder of this paper is organized as follows.
In Section\,\ref{section:High-energy photons} we discuss
possible common sources of gravitational waves and high-energy
electromagnetic radiation (gamma- and X-rays).
In Section\,\ref{section:Medium-energy photons} we highlight the
possible scientific impact of joint observations with infrared,
optical and ultraviolet astronomy.
In Section\,\ref{section:Low-energy photons} we outline the
benefits of coordinated science with radio astronomy.
Section\,\ref{section:Neutrinos} provides insight on the
promise of multimessenger observations with cosmic neutrinos.
Finally, in Section
%s\,\ref{section:Other_Messengers} and
\,\ref{section:Discussion} we consider some additional paths
and benefits of cross-disciplinary studies involving gravitational waves.

\section{High-energy photons} \label{section:High-energy photons}

We begin with an overview of systems that are potentially sources
of both gravitational waves and gamma- or X-ray photons: gamma-ray
bursts (GRBs), soft gamma repeaters (SGRs), ultra-luminous X-ray sources (ULXs),
and micro-quasar flares.  Most of these systems are also
likely candidates for strong neutrino emission; as such we will revisit them 
in Section~\ref{section:Neutrinos}. As well as addressing key
physical questions about
their origin and specific emission mechanisms, multimessenger observations of these
systems may allow us to harness their potential as probes of stellar astrophysics,
galaxy formation, and cosmology.

It is difficult to predict today which instruments will be observing the
electromagnetic spectrum at high energies during the ET era, and in particular
whether an all-sky burst survey (primarily needed to monitor the sources
described in this section) will be on-going. The currently active space missions
Swift, INTEGRAL and Fermi are expected to cease operation before then.
A number of future X- and gamma-ray satellites -- including ASTROSAT (India),
MAXI\footnote{This detector has been installed on the International Space Station during the summer 2009.} 
(Japan) and SVOM (China/France) -- are close to the launch pad but it is
unclear yet whether any of them will still be in operation by 2020. 
Three of the promising missions on the drawing board are the Energetic X-ray 
Imaging Survey Telescope (EXIST)\,\cite{2009AIPC.1133...18G}, 
the International X-ray Observatory (IXO)\,\cite{IXO_WEB_NASA} and 
the Xenia/EDGE mission \cite{xenia}.
The EXIST mission concept is specifically 
optimized for study of high-z GRBs. EXIST promises the detection of $\sim$600 
GRBs/yr and it intends to carry out a rich program targeting transient-source 
science\,\cite{2008ApJ...673.1225B,2006AIPC..836..631G}.

In order to fully exploit the scientific benefits offered by joint observations
with ET, it will be essential to have high-energy satellites operating
during the ET era.

\subsection{Gamma-Ray Bursts}
\label{section:gamma_ray_bursts}

Gamma-ray bursts are extraordinarily luminous flashes of gamma rays
which occur approximately once per day and are isotropically distributed
over the sky; see e.g.~\cite{meszaros_rev} and references therein.
Some GRBs show variability on time scales as short as a
millisecond, indicating that the sources are very compact.
Host galaxies have been identified and their redshifts measured for
more than 100 bursts, demonstrating that GRBs are of extra-galactic origin.

GRBs are grouped into two broad classes by their characteristic
duration and spectral
hardness~\cite{2006Natur.444.1044G,grbs:general:discovery:multiple-classes}.
The progenitors of most short hard bursts (SHBs, with duration $\lesssim$ 2 s 
and hard spectra)
are widely thought to be mergers of neutron star binaries or neutron
star-black hole binaries \cite{Nakar06}.
Long GRBs ($\gtrsim$ 2 s, with soft spectra) are definitively associated
with core-collapse supernovae, particularly type Ic supernovae \cite{Woosley2006}.
Both scenarios are thought to result in the formation of a solar-mass
black hole with a massive ($\sim0.1-1\Msun$) accretion disk.
The gamma rays are thought to be produced by internal shocks in a jet fed by the
accretion disk and powered by its gravitational potential energy or
by the spin of the black hole. 

The LIGO and Virgo detectors have placed upper limits on the strength 
of GWs associated with many individual GRBs 
\cite{GRB030329,070201,S2S3S4,2009arXiv0908.3824L,Ac_etal:08}.  
The most recent LIGO-Virgo search \cite{2009arXiv0908.3824L} placed lower limits 
of 5--20 Mpc on the distance to the GRBs studied, assuming isotropic 
emission of 0.01 $\Msun c^2$ at the network's most sensitive frequency, 150 Hz. 
An analysis of GRB070201, a SHB with sky position error box overlapping M31, 
ruled out the hypothesis that this burst was due to a binary progenitor in M31 
at $>$99\% confidence \cite{070201}.

The binary coalescence leading to a SHB will produce copious amounts of
gravitational radiation that will be easily detectable by ET to $z\sim2-4$. See the article
by van den Broeck {\em et al.\/} in this issue for a discussion of
how joint GW and gamma-ray observations of these systems can be used to
measure cosmological parameters. Specific multimessenger issues relevant
to these cosmological applications of short GRBs are discussed in Section 3.2
below.

For the remainder of this section we focus on long GRBs.  The rate density of observed
long GRBs is estimated at 0.5 Gpc$^{-3}$ yr$^{-1}$
\cite{Sokolov01,Schmidt01,Le:2006pt},
so the typical distance to the closest GRB observed in a year is
$\sim1$ Gpc.  Their progenitors are thought
to be Wolf-Rayet stars -- very massive stars ($>25\Msun$, with
helium cores $>10\Msun$) that have lost their hydrogen mantle.
In the collapsar scenario \cite{collapsar,Wo:93}, the core collapses to form
a $\sim3\Msun$ black hole with a $\gtrsim1\Msun$ accretion disk.
% Numbers 3 and 1 from Piro and Pfahl 2007.  Accretion at $\ge0.1\Msun/s$.
%
% MOTIVATION FOR STUDY:\\
% However, the physics of the system are not understood in detail
% (amount of rotation, role of metallicity and magnetic fields).
It is not yet known which of various proposed mechanisms
is responsible for converting
the disk binding energy or black hole rotational energy into the jets:
neutrino annihilation; magnetic instabilities in the disk; or 
magnetohydrodynamic extraction of the rotational energy.
Other details of the collapsar scenario are also uncertain;
for example, it is possible that the collapsar leads to a
black hole only after fall-back accretion (in which case the
energy source must be magnetohydrodynamic as the neutrino annihilation is too
inefficient at the low accretion rates in this scenario).
In another variant, the supranova model \cite{ViSt:98}, the core
collapse produces a hypermassive neutron star supported by rotation,
which later collapses to a black hole after spinning down due to
dipole radiation.
%Association between GRBs and SN implies delay might be limited to
%a few seconds; if formed, deformation would lead to GW emission.
%
% Yet another possibility:
Other groups have put forward models in which the GRB marks the
birth of a magnetar rather than complete collapse to a black hole
\cite{2000ApJ...537..810W,2001MNRAS.321..177L,2002A&A...391.1141D,Lyutikov:2003ih}.
Since the gamma-ray emission and afterglow are produced at large
distances ($\gtrsim10^{13}$cm) from the central engine, they provide
only indirect evidence for the nature of that engine.  By contrast,
gravitational waves should be produced in the immediate vicinity of
the central engine, offering a direct probe of its physics.

The collapsar scenario requires a rapidly rotating stellar core, so that the
disk is centrifugally supported and able to supply the jet. This rapid rotation
may lead to non-axisymmetric instabilities, such as the fragmentation of the
collapsing core or the development of clumps in the accretion disk.

For example, Davies {\em et al.} \cite{2002ApJ...579L..63D}
suggest that fragmentation is generic,
with a minimum lump size of $\sim0.2 \Msun$. %(smaller lumps decompress explosively).
In this case we may see inspiral-like GW signals for
which the combined component masses are $\sim1.4\Msun$; such signals will be
observable with ET to luminosity distances of order 1 Gpc.
Piro and Pfahl \cite{2007ApJ...658.1173P} argue that gravitational instability coupled
with cooling by helium photodisintegration will produce
$\sim0.1-1\Msun$ neutron star fragments with a lifetime of $\sim1$ s.
For a source at 100 Mpc they estimate the SNR for the advanced LIGO detectors to lie
in the range $1 - 10$, depending on the unknown viscous timescale in the disk which
determines the frequency of transition between viscosity-dominated evolution
and gravitational-radiation dominated evolution.
Scaling to ET sensitivity gives a detection rate of a few times $10^{-3}$yr$^{-1}$
to $\sim2$ yr$^{-1}$ for large viscous timescales, which are expected
for thin disks when neutrino cooling is efficient.
In this case the viscous timescale would be measurable from the peak
of the GW spectrum.

%masses: 1.4-1.4:
%  L2 SNR at 100Mpc = 14.5325
%  L2 Dist at SNR-8 = 181.6559
%  ET SNR at 100Mpc = 220.1486
%  ET Dist at SNR-8 = 2751.8577
%
%masses: 1.2-0.2:
%  L2 SNR at 100Mpc = 4.201
%  L2 Dist at SNR-8 = 52.512
%  ET SNR at 100Mpc = 69.2173
%  ET Dist at SNR-8 = 865.2168
%
%masses: 0.7-0.7:
%  L2 SNR at 100Mpc = 6.9272
%  L2 Dist at SNR-8 = 86.5895
%  ET SNR at 100Mpc = 108.6868
%  ET Dist at SNR-8 = 1358.5855
%
%masses: 0.2-0.2:
%  L2 SNR at 100Mpc = 1.3114
%  L2 Dist at SNR-8 = 16.3919
%  ET SNR at 100Mpc = 22.1238
%  ET Dist at SNR-8 = 276.5471
%
%masses: 0.1-0.1:
%  L2 SNR at 100Mpc = 0.29774
%  L2 Dist at SNR-8 = 3.7217
%  ET SNR at 100Mpc = 6.6126
%  ET Dist at SNR-8 = 82.6581

Alternatively, the suspended accretion model of van Putten
{\em et al.\/}~\cite{vanPutten}, in which the torus is supported
by the residual magnetic field of the star, predicts the development of
strong non-axisymmetries and copious GW emission in a relatively narrow band,
$E_\mathrm{GW} \simeq 0.2\Msun$ at a typical frequency of 500 Hz.
From (\ref{eq:DL}) such emission would be detectable by ET to a distance
of approximately 0.5 Gpc.%, within which the GRB rate is $\simeq2$yr$^{-1}$.

%bar modes: Fryer, Holz, and Hughes 2002 ApJ 565: most optimistic scenario
%bar mode detectable to ~10 Mpc by advLIGO for SN of 15 $\Msun$ ... cross check with Ott and move to SN section, mention briefly here?

%We see that ET limits on GW emission for a source at 150 Mpc will be around $10^{-5}M_\odot$ at 50 Hz; this is sufficient to detect or rule out processes like fragmentation of the collapsar core or bar mode instabilities, and perhaps some less-exotic processes seen in SN simulations (and certainly the van Putten model).  150 Mpc is in turn the distance within which we expect one-several low-luminosity GRBs per year.  {\bf Should probably move the introduction to low-luminosity GRBs from the HEN section to here, assuming this section appears earlier.}

% FREQUENCY:\\
% At most 1\% of core-collapse SN can be associated with GRBs \cite{GalYam_etal:06}.
% Those associated with GRBs are likely the most massive and most
% rapidly rotating stars, possibly favoured in regions of low metallicity.

%\begin{figure}
%\includegraphics[width=0.75\columnwidth]{distance_vs_freq}
%\caption{(duck!)}
%\label{fig:GRB}
%\end{figure}

Long GRBs appear to include a sub-class known as ``low-luminosity GRBs,''
which are associated with particularly energetic core-collapse supernovae.
Examples of these objects include GRB980425 / SN1998bw
\cite{Galama:1998ea,1998Natur.395..663K}, GRB031203 / SN2003lw
\cite{2004ApJ...609L...5M,2004Natur.430..648S}, and GRB060218 / SN2006aj
\cite{Campana:2006qe,2006ApJ...645L.113C,2006Natur.442.1011P,2006Natur.442.1014S}.
As these events are less luminous than typical long GRBs, they are often
discovered at smaller distances, for example: SN1998bw at redshift $z = 0.0085$, less than
40~Mpc from Earth; SN2003lw at $z = 0.105$; SN2006aj at $z =0.033$.
Recent studies \cite{Liang2007,2006Natur.442.1014S,2007MNRAS.382L..21C} indicate the
local rate density of under-luminous long GRBs
may be as much as $10^3$ times that of the high-luminosity population -- with
the closest such GRB observed, in one year of operation, lying at a typical distance of
only 100 Mpc.
This is an encouraging prospect for their detection in GWs, particularly
given the consensus that these events are the extreme end of a continuum of
events with the same underlying physical model, rather than physically
distinct progenitors \cite{Woosley2006}.
In the same way, X-ray flashes (XRFs) and X-ray rich (XRR) GRBs \cite{He_etal:01}
are observationally similar to ordinary long GRBs,
% with softer spectra and similar energy release per solid angle.  These
and could also be produced by the same underlying progenitors.
% as GRBs, simply observed slightly off-axis or other factors (see refs in \cite{Woosley2006}).
Again, ET observations may confirm or refute this conjecture.

%Central engine directly observable in GWs.
%How long does the central engine operate?  Is the power at late times episodic or continuous?

%Merger phase of compact body coalescence and collapsar:
%$E_{GW}^{iso} \sim 5\times10^{-2} \Msun \sim 5\times10^{52}$ erg.

%Must cite work of Kobayashi and {M{\'e}sz{\'a}ros} \cite{2003ApJ...589..861K}
%on GW emission by GRBs.

\subsection{Soft gamma-ray repeaters}

%(MeV emission)

%{\em INSERT: Grab results from LSC PRL \& ApJL papers, also QPO PRD,
%scale to ET sensitivity curve.  Should get Figure~\ref{fig:SGR}, or
%something very close.  Assume intrinsic isotropic energies of GW sources.
%For Soft Gamma Repeaters, upper limit on EGWiso ~1046 erg (see refs in
%Astrophys. J. 681 (2008) 1419 and arXiv:0808.2050,
%particularly Ioka), comparable to EM emission.
%How about more conservative models too?}

Soft gamma-ray repeaters are systems that emit brief bursts of soft gamma rays
and X-rays at irregular intervals \cite{Mereghetti:2008je,wod}.
These bursts have typical durations of $\sim0.1$ s and luminosities up
to $10^{42}$ erg s$^{-1}$.  Three of the five known SGRs have also been observed
to emit rare ``giant flares'' with luminosities up to $10^{47}$ erg s$^{-1}$ 
and total gamma-ray emissions up to $10^{46}$ erg.

According to the magnetar model, SGRs are galactic neutron stars with
magnetic fields of $\sim10^{15}$ G.  Flares occur when the solid crust
cracks due to deformations induced by the magnetic field
\cite{thompson95,schwartz05,Horowitz:2009ya}.
% Giant flares are generated by catastrophic re-arrangement of the crust and magnetic field.
This cracking may excite the star's nonradial modes,
particularly $f$ modes \cite{andersson97,pacheco98,ioka01},
producing GWs \cite{Horvath2005,owen05}.
% Emission would also occur if SGRs are solid quark stars.
%The strength of the GW emission depends on unknown coupling
%strengths, mode excitations.
The most optimistic estimates for the energy reservoir available
in a giant flare is $10^{49}$ erg \cite{ioka01}; a more recent
analysis \cite{corsi09} indicates that a more realistic
limit is between $10^{45}$ erg (about the same as the total EM
emission) and $10^{47}$ erg.  However, the efficiency of
conversion of this energy to GWs is unknown.
%Normal NS can store $\sim10^{44}$ erg in crustal elastic energy [PRD],
%quark star perhaps $10^{46}$ erg [PRD].
%Ioka: Grav pot energy much larger than magn field; could tap into.
%The most efficient GW emission is expected to be in $f$ modes,
%at frequencies of 1-3 kHz \cite{Andersson_and_refs_in_ApJL}.

LIGO has placed upper limits \cite{SGRpaper,abbott:211102,Collaboration:2009zd} on
GW emission by SGRs in the range $10^{45}-10^{51}$ erg, depending on frequency,
and assuming a nominal source distance of 10 kpc.  Typical LIGO limits in the
1-3 kHz range, expected for $f$ modes, are $10^{49}-10^{50}$~erg.

\begin{figure}
\begin{center}
\includegraphics[width=0.75\columnwidth]{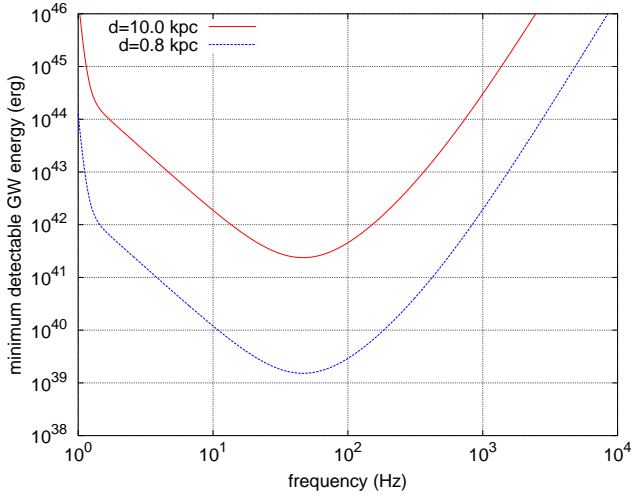}
\caption{Minimum energy in gravitational waves detectable by ET
as a function of frequency for an SGR source at 10 kpc (solid line)
and 0.8 kpc (dashed line).  This limit assumes isotropic and
narrowband GW emission.}
\label{fig:SGR}
\end{center}
\end{figure}

Figure~\ref{fig:SGR} shows the sensitivity of ET to an SGR source
at a distance of 10 kpc (a typical galactic distance) and 0.8 kpc
(the estimated distance to SGR 0501+4516 \cite{2008GCN..8113....1B,2008GCN..8149....1G,Leahy2007}).
At $f$-mode frequencies ET will be sensitive to GW emissions
as low as $10^{42}$ -- $10^{44}$ erg at 0.8 kpc, or about 0.01\% to 1\% of the energy content
in the EM emission in a giant flare.  In the region of $20 - 100$ Hz,
ET will be able to probe emissions as low as $10^{39}$ erg, i.e. as
little as $10^{-7}$ of the total energy budget.

%Pulsation in the light curve tail reveal the NS spin frequency.
%Quasiperiodic oscillations have been seen in the light curves of SGR 1900+14 and SGR 1806-20.

\subsection{Ultra Luminous X-ray binaries (ULXs)}

Stellar-mass black holes (${\rm M}<20\Msun$) have an Eddington luminosity of $10^{39}$ erg/s.
This fixes an upper limit on the luminosity for ``normal'' X-ray binaries.
However, several objects (the nearest being in M33) have observed luminosities (i.e. inferred
from their measured fluxes and assuming isotropic emission) in the range $10^{39}$ to $10^{41}$ erg/s.  Three scenarios have been proposed to explain this apparent anomaly:
\begin{itemize}
\item ULXs contain a stellar-mass black hole but the X-ray emission from the surrounding accretion disk is beamed, implying a significantly lower (sub-Eddington) luminosity;
\item ULXs contain a stellar-mass black hole and the X-ray emission from the accretion disk {\em is\/} isotropic, thus implying a super-Eddington luminosity;
\item ULXs contain an intermediate-mass black hole (IMBH), with ${\rm M}>100\Msun$, for which the Eddington luminosity is compatible with the luminosities observed for these systems.
\end{itemize}

While the third scenario offers an intriguing explanation of the ULX phenomenon, the
existence of IMBHs is not well established \footnote{In \cite{Strohmayer:2009be}, the authors claim to have observed one associated with the ULX NGC 5408 X-1.}. These objects have been proposed as light `seeds' of the massive black holes found in the centres of galaxies; they are thought to form
at high redshift as the end product of the first generation of stars.
Should these IMBHs exist, the GW signatures of their mergers will likely fall between the sensitivity bands of 2nd generation ground-based detectors and LISA. However, these merger events would be bright GW sources for ET, visible
to $z \sim 10$. Recently Sesana {\em et al.\/}~\cite{2009ApJ...698L.129S}
computed the number of IMBH
merger events, for a  variety of black hole `seed' formation models, that would be detected
in three years of operation by a single ET with 10 km arms. They found that for almost all
models ET could expect to observe a few, to a few tens of mergers per year -- with the black hole masses measurable from analysis of the waveform during the inspiral phase. 
While not resulting strictly speaking from a multimessenger approach,
observations with ET may, therefore, help to discriminate between the proposed ULX scenarios
since they should confirm or refute the existence of IMBHs.

\subsection{Microquasar flares}

%(keV emission)

Microquasars \cite{Mirabel:1999fy} are radio-emitting X-ray binaries.
Their name is motivated by their observational similarities to quasars,
particularly their strong, variable
radio emission and the presence of radio jets.  The jets are powered by
accretion from a normal companion star onto a central object which may be a
neutron star or a solar-mass black hole.  The accretion disk is very
luminous in the optical and X-ray regimes.

%Microquasars are very important for the study of relativistic jets. The jets are
%formed close to the compact object, and timescales near the compact object are
%proportional to the mass of the compact object. Therefore, ordinary quasars take
%centuries to go through variations a microquasar experiences in one day.
%
% e.g.: SS 433, in which atomic emission lines are visible from both jets
% GRS 1915+105, with an especially high jet velocity
% the very bright Cygnus X-1;
%and the microquasar LS I +61 303, which has been discovered to emit VHE gamma rays.

Microquasars exhibit X-ray and radio/IR flares which may be
explained by ejection of ``blobs'' of accreting matter.  The X-rays originate
from the inner accretion disk, while the radio/IR emission is due to the
ejection of ultra-relativistic blobs of plasma (ballistic motion).  In this
``cannonball model'' one can expect the emission of a GW burst with memory
from the microquasar, with typical strain amplitude
\cite{PhysRevD.64.064018,Pradier:2008uj}
\begin{equation}
h \sim 10^{-22} \frac{\Gamma}{10}\frac{m}{10^{-7}\Msun}\frac{1\mathrm{kpc}}{D} \, .
\end{equation}
Here $D$ is the distance to the microquasar, $m$ is the mass of the blob,
and $\Gamma$ is the Lorentz factor; the nominal value of $m$ is
of order of the mass of the Moon.
The corresponding energy in GWs is of order
\begin{equation}
E_\mathrm{GW} \sim \frac{c^3}{8 G} \frac{h^2 D^2}{T},
\end{equation}
where $T$ is the duration of the burst.

The acceleration time of the blob determines the duration of the burst
and hence its typical frequency.  This may range from $T \sim 10^{-5} s$
(the free-fall time for a solar-mass compact object) up to minutes.
%The free-fall time into the central object provides an
%estimate of the acceleration time. The burst duration thus scales with
%$T=R^{3/2}/\sqrt{G M}$ where $R$ and $M$ are the radius and mass of the central
%object.
An acceleration timescale of $\sim1$ ms would place most of the radiated
energy $E_\mathrm{GW} \sim 10^{-15}~\Msun = 2\times10^{39}$ erg at
frequencies around 100 Hz.  From equation (\ref{eq:DL}), this would
yield a typical signal-to-noise ratio of $\rho\sim6$ at $1$ kpc;
such a GW burst might be marginally detectable by ET.
The observation of such a GW pulse coincident in time and direction with a
microquasar flare would confirm that the jet is relativistic and would possibly
provide an independent measurement of the mass of the ejecta.

%ET may be able to detect or constrain the cannonball model for micro-quasar flares.

% blob mass of the moon
% h \sim 8 \times 10^{-24} \frac{\Gamma}{10}\frac{m}{4 \times 10^{-8}\Msun}\frac{5\mathrm{kpc}}{d}
% Egw=2e-14 msol c^2 = 5.0e+40 erg, that's much closer to what ET can detect

% The typical frequency of the burst is inversely proportional to the
% acceleration time of the blob, which may be estimated as the
% gravitational time scale
% \begin{equation}
% t_{\mathrm{acc}} \sim \frac{10^5\mathrm{Hz}}{10\Msun}
% \end{equation}
% and the size of the accretion disk
% \begin{equation}
% r_{\mathrm{disk}} \sim 10^2\mathrm{Hz} \frac{r}{10^3\mathrm{km}}
% \end{equation}

% Refs: Mirabel and Rodriguez, 1999, Pradier, arXiv:0807.2562v1 Segalis and Ori, 2001

\section{Medium-energy photons} \label{section:Medium-energy photons}

%\textcolor{red}{Need a few words of introduction: Begin with optical supernovae; example of triggered search similar to what has been discussed in previous section
%for GRBs and SGRs.  Follow with discussion of opportunities and challenges of
%using GW observations to trigger EM follow-ups.}

As discussed in Sections 1 and 2,
the well-established history of multimessenger astronomy involving GWs
has focussed mainly on high-energy EM (i.e., gamma- and X-ray)
observations being used to trigger searches in GW data for signals
associated with GRBs and SGRs.
%postponed to neutrino section:
%A similar procedure is also envisaged in the event
%of a low-energy neutrino (SNEWS) alert for a nearby supernova.
By contrast
efforts are only now beginning on joint observations between GWs and the
medium and low-energy EM spectrum (optical, radio, etc.), and in this context
we can also envision an alternative procedure, where EM observations are
triggered by GW detections.

While the range of possibilities is great when considering the science
enabled by joint optical and gravitational wave observation, here we will only
discuss two of the best surveyed and understood topics: optically detected
core-collapse supernovae and short GRBs.  Both of these are potentially
interesting as GW sources.

For optically detected extragalactic supernovae,
%we have an example of an
%EM-triggered search similar to those discussed in Section 2. Here
the GW arrival time must be predicted from data derived from an early optical detection.
This results in a large uncertainty (of order several hours) on the predicted
arrival time, which makes the GW data analysis task challenging. However,
information on the direction -- and to a lesser extent the distance -- of an
optically detected supernova is usually rather precise, which at least
permits the use of directional GW search analysis methods. The large
number of extragalactic supernova discoveries expected in the future,
from e.g. ground-based optical telescopes such as PanSTARRS and LSST (see below),
makes this line of analysis potentially very fruitful -- even though its
theoretical motivation is still evolving.

Targetted observations in the optical and infrared of short GRBs shall
help us measure the redshift of these coalescing compact binary systems.
The coincident gravitational wave amplitude, frequency and frequency
derivative during inspiral and coalescence phases can yield a precise
estimate of their luminosity distance. The combination of the two
could in turn provide a completely independent way to calibrate the
cosmological distance scale and constrain cosmological models.

\subsection{Optically Detected Core-Collapse Supernovae} \label{subsection:Optical_Supernovae}

Core-collapse supernovae have long been considered as one of the most
interesting targets for gravitational-wave observations.  Many different
physical phenomena during and after the collapse have been studied in the
context of GW emission; unfortunately, most studies indicate that even
the second-generation LIGO detectors will only be able to detect GWs from
supernovae in our Galaxy.  Since the galactic supernova rate is approximately
0.02 yr$^{-1}$, the GW observation of a SN will have to rely on
luck -- until ET.

\begin{figure}
\includegraphics[width=1.00\columnwidth]{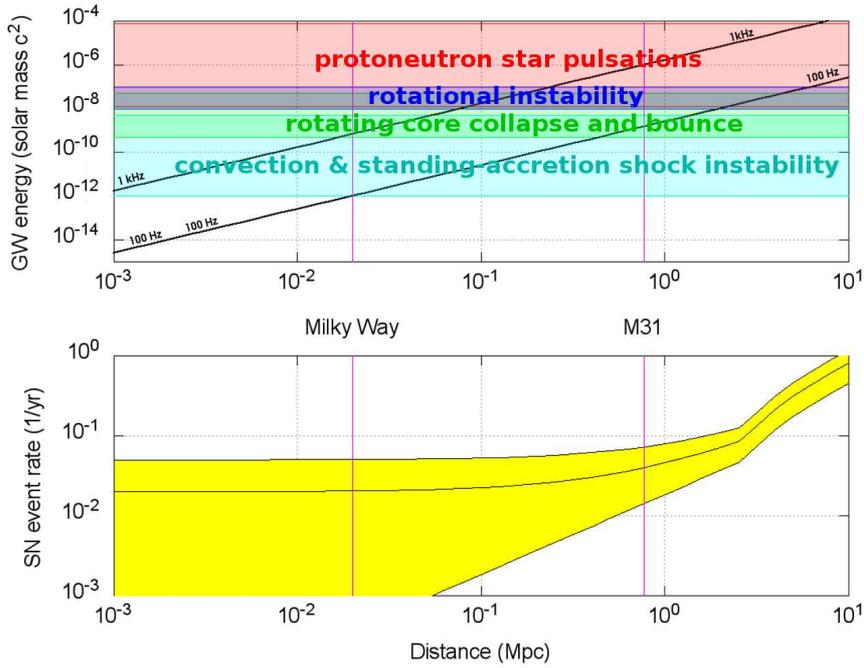}
\caption{\label{fig:Supernovae_Rate} The upper plot displays the minimum GW
  energy that a supernova core collapse is required to radiate in order to be
  detectable by the Einstein telescope. We give two estimates obtained from
  Eq. (\ref{eq:DL}) assuming a GW signature with a low frequency content $f=100$
  Hz and high frequency content $f=1$ kHz. This quantity is given as a function
  of the source distance.  We also indicate the expected range of radiated GW
  energy for several processes \cite{Ott2008}. The lower plot shows an estimate
  of the cumulative event rate (with error bars) obtained from the star
  formation rate computed over a catalog of nearby galaxies
  \cite{2005PhRvL..95f1103A}.}
\label{fig:SN}
\end{figure}

The solid line on the upper panel of Figure\,\ref{fig:Supernovae_Rate} shows the
GW energy that a supernova core collapse is required to radiate, in order to
be detectable by the Einstein Telescope. Also shown as horizontal bands are
the predicted ranges of GW energy for several theoretical SN mechanisms.
The lower panel of Figure\,\ref{fig:Supernovae_Rate} shows the expected
cumulative event rate as a function of distance. While the upper plot
indicates that a significant fraction of the modelled waveforms should
be visible to ET from the distance of M31 ($\sim$770 kpc), the lower
plot shows that the event rate at that distance is still low -- likely
requiring decade(s) of observation to secure a single supernova detection.
Nevertheless, we note from the upper plot that {\em some\/}
mechanisms are predicted to be sufficiently energetic to be detectable by ET
at a distance of up to 10 Mpc, where the SN event rate reaches a value of order
unity. Therefore, extended observations by ET should be able to detect
or place strong constraints on the role of the more energetic theoretical
processes shown in Figure\,\ref{fig:Supernovae_Rate}.

Ott\,\cite{Ott2008} provides an in-depth and comprehensive review of the
state-of-the-art understanding of the GW signatures of core-collapse supernovae.
Section\,\ref{section:Neutrinos} of this paper discusses supernovae in more
detail and outlines the possible benefits of observing them jointly in
low-energy neutrinos and gravitational waves.

%in red, protoneutron star pulsations $E_{GW} \sim 1.2 \times 10^{-8}$ to $8.0
%\times 10^{-5} M_{\odot} c^2$; in green, rotating core collapse and bounce
%$E_{GW} \sim 5 \times 10^{-10}$ to $5.0 \times 10^{-8} M_{\odot} c^2$; in blue,
%rotational instability $E_{GW} \sim 10^{-8}$ to $10^{-7} M_{\odot} c^2$; in
%cyan, convection and standing-accretion shock instability $E_{GW} \sim 10^{-12}$
%to $5.0 \times 10^{-9} M_{\odot} c^2$.

\subsection{Gravitational wave standard sirens}

Perhaps the clearest role for multimessenger observations in the optical and infrared lies in the exploitation of very deep imaging and spectroscopic data to identify the host galaxy, and thence
measure the redshift, of coalescing compact binary systems --
so-called ``gravitational wave standard sirens''.

In the past few years there has been growing interest in the potential future use of these systems as high-precision cosmological distance indicators, since measurement of their gravitational wave amplitude, frequency and frequency derivative during inspiral and coalescence can yield a precise estimate of their luminosity distance~\cite{1986Natur.323..310S}. While much attention has been focussed on binary supermassive black hole mergers that will be a major observational target of the LISA satellite~\cite{2005ApJ...629...15H,2009CQGra..26i4027A}, recently
Nissanke {\em et al.\/}~\cite{2009arXiv0904.1017N}, extending earlier work by
Dalal {\em et al.\/}~\cite{2006PhRvD..74f3006D}, have investigated the prospects for detecting binary neutron star mergers with a network of advanced ground-based detectors. They showed that the source's luminosity distance could be determined to an accuracy of better than 30\%, out to a distance of 600 Mpc. Thus, gravitational wave standard sirens could provide a completely independent way to calibrate the cosmological distance scale and constrain cosmological models via their luminosity distance redshift relation -- in a manner complementary to other, electromagnetic, cosmological probes~\cite{2009astro2010S.130H}.

Crucial to their usefulness for cosmology, however, is the inference of
each source's redshift, which is not possible from the gravitational wave data alone. This {\em may\/} not require explicit association of the source with a single, host galaxy: MacLeod and Hogan~\cite{2008PhRvD..77d3512M} have developed a interesting method for obtaining unbiased estimates of standard siren redshifts by exploiting information about the clustering of galaxies in each source's 3-D error box. However definitive identification of the source's electromagnetic counterpart, and thus measurement of an accurate sky location, is in any case highly desirable since it will also significantly improve the determination of the source's intrinsic parameters -- including its luminosity distance -- by breaking parameter degeneracies and avoiding  the need to marginalise over sky position
~\cite{2009CQGra..26i4027A,2003CQGra..20S..65H}.

Recently Sathyaprakash {\em et al.\/}~\cite{2009arXiv0906.4151S} 
(see also van den Broeck {\em et al.\/}, this volume) have considered 
the prospects for using the Einstein Telescope as a precision tool for 
cosmology. The annual rate of coalescence for binary neutron star and 
neutron star-black hole systems within the volume observable by ET 
%(i.e. out to $z \sim 8$) % -- horizon distance, not average distance (z_avg~4).
is expected to be several $\times 10^5$. As discussed in the previous 
section, these systems are believed to be the progenitors of short-hard 
gamma ray bursts. Sathyaprakash et al. assumed that over a three-year 
period of ET operation the electromagnetic counterparts of about 1,000 
standard sirens, distributed with constant comoving number density over 
the redshift range $0 \leq z \leq 2$, would be detected optically, and 
their redshifts and sky positions measured. The authors showed that, 
with these observations, the dark energy equation of state parameter $w$ 
could be measured to a precision of 15\% ($1-\sigma$ error) -- assuming 
also that the effects of weak lensing could be corrected (see below and~\cite{2009arXiv0907.3635S} for further discussion). In fact if $w$ were 
the {\em only\/} parameter to be fitted (e.g. if the dimensionless 
density parameter were assumed known, from other cosmological 
observations, and the Universe were assumed to be flat) then $w$ could 
be measured to a precision of about 1\% ($1-\sigma$ error).

Is identifying the optical counterpart and host galaxy of GW sirens observed
by ET out to $z = 2$ a realistic prospect? As pointed out in
Bloom {\em et al.\/}~\cite{2009arXiv0902.1527B}, short-hard GRBs are now known
to produce faint optical and infrared afterglows, detectable for a few days
with current and planned future instrumentation.
Perley {\em et al.\/}~\cite{2009ApJ...696.1871P} compare light curve models
for a $^{56}$Ni-powered ``mini-SN''~\cite{1998ApJ...507L..59L} with optical
observations of the transient associated with the short-hard GRB 080503;
these data may represent the first electromagnetic signature of a binary
neutron star inspiral. The optical observations of the afterglow were
extremely faint, however -- never exceeding 25th magnitude -- and no
coincident host galaxy brighter than 28.5 mag. was found with HST.

On the other hand Berger {\em et al.\/}~\cite{2007ApJ...664.1000B} presented
optical observations of nine short-hard GRBs, obtained with Gemini, Magellan,
and the Hubble Space Telescope.  They identified candidate host galaxies from
optical and X-ray afterglow observations, and measured spectroscopic redshifts
in the approximate range $0.4 \leq z \leq 1.1$ for four of them. For eight of
the nine GRBs in this sample the most probable host galaxy had an $R$-band
magnitude in the range $R \sim 23 - 26.5$ mag. This is certainly faint by the
standards of current instrumentation and, together with the lack of an
HST-detected host galaxy for GRB 080503, underlines the current observational
challenge of determining the redshift of more distant sirens.

However it is expected that by 2020 the available ground-based optical and infrared facilities will include the proposed European Extremely Large Telescope (ELT)~\cite{eelt} and the Thirty Meter Telescope~\cite{tmt} -- both of which could begin operations as early as 2018. These instruments will target observations of `First Light' galaxies at high redshift, and should be capable of obtaining high quality spectra, even at $z \sim 6$ with modest $\sim 1$ hour integration times, from galaxies significantly fainter than $L^*$, which is a parameter (representing the luminosity of a typical galaxy) of the Schechter function widely used to model the distribution of galaxy luminosities.  The task of determining standard siren redshifts out to $z\sim2$ from imaging and spectroscopy of their host galaxies should, therefore, be straightforward\footnote{Indeed this task should
also be possible in the radio, assuming that the SKA begins operation on a similar timescale to ET. See also Section 4.}.

Of course the above remarks make an important assumption: that an electromagnetic counterpart of the short-hard GRB is identified, in X-rays, UV or optical, from a wide-spectrum monitoring satellite such
as SWIFT. Five of the short-hard GRBs considered in Berger {\em et al.\/}, for example, had no observed
optical afterglow but had X-ray positions measured to within a radius of 6 arcseconds -- sufficient to
identify the probable host galaxy with good confidence. In the absence of this precise directional information the task of determining the siren redshift is rendered significantly more difficult: in addition to the source parameter degeneracies discussed above the issue of source confusion becomes very serious, as we now briefly consider.

The angular size of the source position error box derived from GW observations alone will be large. Recent analyses (see e.g.~\cite{2006PhRvD..74h2004C,2008JPhCS.122a2001B,Fa:09}) suggest that a network of second generation detectors will locate NS binaries to within a field of about 10 square degrees, and
a similar angular precision could be expected from a network of ETs. A realistic future observational strategy for second and third generation detectors may, therefore, require using deep and wide optical/infrared survey data to search for siren host galaxies. Crucial for this purpose will be facilities such as the Large Synoptic Survey Telescope (LSST,~\cite{2008arXiv0805.2366I}): this is an optical imaging survey instrument with a $\sim 10$ square degree field of view, which will map 20,000 square degrees in 6 photometric bands to a depth of $~ 0.3 \mu$Jy.  LSST will potentially be a very powerful tool for multimessenger astronomy with ET -- both via a `rapid response' mode that would slew to the putative GW source location to search for transient optical signatures, and via the use of archival `survey mode' data to identify candidate host galaxies within the GW source error box.

The potential number of such candidate hosts can be gauged by considering, for example the Hubble Ultra Deep Field which contained more than 8000 galaxy detections within an area of 12 square arcmin. Scaling this number density to an area of 10 square degrees would yield 24 million galaxies!  Of course one is searching amongst those galaxies for a specific optical transient signature, but with so many galaxies in the error box several such transients may be observed. This underlines the importance of developing a much better theoretical understanding of the electromagnetic signatures of short-hard GRBs in the decade or so before ET begins operation.  Without a clear pre- and/or post-merger electromagnetic signature to narrow the search, identifying the siren's host galaxy becomes like searching for a small needle in a very large haystack. Efforts to better understand the electromagnetic signatures of supermassive black hole mergers for LISA are already well underway (see, for example,~\cite{2006ApJ...637...27K,2008PhRvL.101d1101K,2009ApJ...700.1952H}) and a similar effort for short-hard GRBs should be undertaken. Of course
one advantage for ET, unlike LISA, is that we can confidently expect the detection of a number of binary inspirals with second generation detectors.
Thus there should be several years' worth of intensive multimessenger observations of standard sirens to draw upon before ET begins operation. These
observations should hopefully provide a powerful incentive to drive forward our theoretical understanding of these systems, and the electromagnetic
signatures that will reveal their presence in the skies probed by ET.

Finally, we should mention another important multimessenger role for deep and wide optical/infrared survey data in the context of gravitational wave standard
sirens. For several years it has been recognised that the performance of sirens as cosmological probes will be significantly undermined at high redshift by the impact of weak gravitational lensing from intervening large-scale structure~\cite{2005ApJ...629...15H,2006PhRvD..74h2004C,2009arXiv0906.4151S}. One possible method to mitigate this problem is to construct
high-resolution maps of the cosmic shear and convergence along the line of sight to a siren, in order to identify -- and attempt to correct for --
the amount of weak lensing (de-)magnification. Recently Shapiro {\em et al.\/} \cite{2009arXiv0907.3635S} have investigated this approach in
detail for putative LISA supermassive black hole binaries, although the methodology would be equally applicable to standard sirens observed with ET. The
authors demonstrate that the weak lensing scatter can indeed be partially corrected so as to reduce the distance error dispersion by up to 50\% for
a source at $z = 2$. However this result assumes the availability of weak lensing magnification maps constructed from a 2-D ELT mosaic image and
a wide-field image from a space survey telescope such as JDEM or EUCLID.  Thus the requirement for a multimessenger approach is also very
strong in this context.

\section{Low-energy photons} \label{section:Low-energy photons}

%\begin{itemize}
%\item Binary mergers (\& short GRBs)
%\end{itemize}

% text from white paper draft

\subsection{Radio Astronomy}

The existence of theoretical models in which various mechanisms give rise to a prompt pulse of radio emission from some putative gravitational wave sources, particularly coalescing compact binaries, motivates joint GW searches with radio telescopes. Moreover, the use of GW detectors as a trigger for
follow-up radio searches could provide a method of detecting faint radio
transients that might otherwise be missed.

Future ground-based radio telescopes, such as LOFAR~\cite{lofar} and the Square Kilometre Array~\cite{ska}, will employ aperture synthesis to allow
the observations to combine a field of view that is a large fraction of the sky with an angular resolution that is a fraction of an arc second. Such telescopes could therefore use digital signal processing to match the telescope field of view with the error circle of a gravitational wave trigger
supplied by a network of ETs. The bandwidth of LOFAR is limited to the UHF (40-240~MHz) by the useable bandwidth of the antennae. For other conventional steerable dish telescopes, with a much smaller
field of view, the task of identifying the radio counterpart of a GW source is more challenging, although these telescopes do at least have the advantage of covering a wide range of radio frequencies across a
frequency band for which the radio counterpart signature is better understood. Hence steerable dish telescopes could be employed for dedicated follow up observations, once the source location had been determined precisely using LOFAR or the SKA. Moreover the use of gravitational waves as the search trigger would at least make optimal use of the wide field of view of GW interferometers to maximize the probability of transient detection in GWs. Furthermore

What would multimessenger astronomers `see' with their radio telescopes?  Models for prompt radio emission from compact binaries generally require that one of the compact objects is a magnetar. In the simplest of these model classes  \cite{lipunov,hansen}, the orbital motion of the binary generates time dependent magnetic fields and consequently induced electric and magnetic fields.  These fields then lead to the emission of radiation, which is predicted to be in the radio band.

A second, larger class of models similarly require a high magnetic field from
one member of the binary; in these models the field either confines or otherwise interacts with a plasma. For example, the unmagnetized companion object can develop surface charge that can then be ejected from the surface of the star and subsequently undergo acceleration as it follows the magnetic field lines of the magnetar. Alternatively, gravitational waves emitted during the inspiral and merger of a binary neutron star system may excite magnetohydrodynamic waves in the plasma (see \cite{moortgat,shapiroMHDGW} for details), which then interact with charged particles in the plasma, inducing coherent radio emission.

There are a series of proposed models for radio afterglows for SHBs that may be
observed seconds to minutes after the burst. One of the models put forward by
Usov and Katz~\cite{usov} predicts that immediately after merger the rotational
energy of the binary system is transferred to a highly magnetized, highly
relativistic particle wind that interacts with the ambient warm gas and as a
result EM radiation is emitted. The main bulk of the radiation is below 1 MHz
but its tail can reach 1--30 MHz. A key prediction of their model is that there
should be an incoherent radio pulse in the frequency range $1-30$ MHz, with a
duration of $1-100$ seconds and a fluence of a few percent of the total
gamma-ray fluence from the source.  The expected time delay for the pulse would
be around $10^4$ s for a source placed at 3.2 Gpc.
Published results on radio afterglows for SHBs 
show only weak signals hours or days after the burst.

The detection of such a radio pulse might well require observations at
lower frequencies than are usual for ground-based radio astronomy. Even
the LOFAR low frequency radio array~\cite{lofar} is not sensitive below about 30 MHz, and the proposed Phase 2 design specification for the Square
Kilometre Array radio telescope~\cite{ska} extends only to 70 MHz. Thus,
as acknowledged by Usov and Katz, radio observations from space may be required to detect the afterglow signatures predicted in their model: these observations would be free from the effects of ionospheric refraction, although interstellar scintillation would still be very strong.

On the other hand, higher frequency radio afterglow signatures -- while harder to predict theoretically at present -- might be accessible from the ground.  In that regard (as was similarly noted in Section 3.2 in the context of optical signatures) before the ET era begins we can confidently expect the
detection of a number of SHBs with second generation GW detectors.  These
should present important opportunities to carry out radio follow-up searches, using e.g. LOFAR, the various SKA precursors and indeed possibly
the SKA itself.  These crucial first GW discovery events will, therefore, hopefully equip us with a much better understanding of their radio
afterglows in the frequency range that will be accessible to ground-based radio telescopes during the ET era.

In fact prospects for detecting the radio {\em precursor\/} of an SHB appear to be better. Hansen and Lyutikov~\cite{hansen} have modelled the electromagnetic signature expected from the magnetospheric interactions of a neutron star binary prior to merger. In view of the lack of a complete theory of pulsar radio emission, they adopted a simple parameterisation based on what is known about pulsars; nevertheless they found that detectable signals were possible in both radio and X-rays, estimating the radio flux density at 400 MHz to be
\begin{equation}
F_\nu \sim 2.1 {\mathrm {mJy}} \frac{\epsilon}{0.1} \left ( \frac{D}{100 {\mathrm{Mpc}}} \right) ^{-2} \, B_{15}^{2/3} \, a_7^{-5/2} ,
\end{equation}
where $\epsilon$ is a dimensionless efficiency factor, $D$ is the
distance to the binary, $B_{15}$ is the magnetic field in units of
$10^{15}$G, and $a_7$ is the orbital semi-major axis of the binary
in units of $10^{7}$cm.
As noted by Hansen and Lyutikov, for an SHB within a few hundred Mpc 
this emission would already in principle be detectable by the larger 
radio telescopes operating today -- although it would lie somewhat 
below the sensitivities of current radio transient searches. If the 
model predictions of these authors are reasonable, therefore, 
pre-merger radio emission from neutron star--neutron star binaries 
should be a straightforward observational 
target for LOFAR and the SKA out to cosmological distances -- at least 
provided the problem of source confusion is overcome and the radio and 
GW emission are both associated with the correct source.

To summarise, all of the above suggests interesting future possibilities 
for joint radio and GW observations of SHBs.  For binary mergers ET will 
provide a very precise time of coalescence (to within milliseconds).  It 
will also provide the redshifted masses and spins of the binary components, 
a rough sky position (to within about $\sim$10 deg$^2$), and an approximate 
luminosity distance (perhaps to a factor of 2).  These data could then be 
used to trigger detailed follow-up radio observations, to
identify the radio transient associated with the GW signal. While the task
of identifying this radio counterpart is clearly challenging, sharing many
of the source confusion issues discussed in Section 3.2, note that we can
determine the dispersion measure for any candidate counterpart. This will 
provide an independent measure of the distance, allowing us to better
predict when the GW emission should have arrived at our detectors -- and
thus hopefully narrowing the search for the `true' radio counterpart.

Most importantly, a robust and reliable identification of the radio
transient associated with a binary merger event could open the door to using the full binary population (of order $10^6$ systems) observed by ET for cosmological measurements, rather than just the small fraction of binaries which are also detected as GRBs. Moreover, determining the redshift of the
host galaxy could also be carried out in the radio -- thus avoiding the need to identify the host galaxy optically and so providing an alternative route
to mapping out the luminosity distance redshift relation for GW sirens.

\section{Neutrinos} \label{section:Neutrinos}

Many of the astrophysical systems observable by ET are also expected to be
strong emitters of neutrinos. Two energy ranges of the neutrino spectrum are of
observational interest: low energies, $E_{\nu}\lesssim 10$ MeV; and high
energies, $E_{\nu} \gtrsim 100$ GeV. (The intermediate range around $100$ MeV is
currently inaccessible to earth-based detectors because it is overwhelmed by
atmospheric neutrinos from air showers). Cosmic neutrinos in these different energy
ranges originate from different astrophysical processes. Consequently in this section
we treat low and high energies separately when compiling a list of potential joint
sources of gravitational waves and neutrinos.

Neutrino observatories have quantitatively similar characteristics -- in terms of FOV,
angular resolution and distance sensitivity -- to GW observatories. Hence, for
many astrophysical sources joint neutrino and GW observations would represent
a ``marriage of equals'' -- although a notable exception would be the case of a
galactic core-collapse supernova, which for current GW and neutrino detector
sensitivities would be a much higher SNR source of low energy neutrinos than of
gravitational waves.

% Neutrino detectors exploit the fact that neutrinos interact with matter with a
% small but non-zero probability, thus producing relativistic (charged) particles
% through the charged-current weak interaction. If such interactions occur in a
% medium that is transparent to visible light, the resulting particles propagate
% at speeds greater than the speed of light in that medium, and thus a flash of
% Cherenkov light is released. The energy range of the neutrino determines the
% size of the detector.

For the low-energy range, detectors are composed of a vessel filled with pure
water or a liquid scintillator. The sensitivity of the neutrino detectors scales
linearly with the mass of liquid. %, which is typically in the kiloton range.
The main detectors\cite{fulgione10:_status} currently are Super-Kamiokande
(Japan) with 50 kilotons of pure water, and LVD (Italy), KamLAND (US/Japan) and
the upcoming SNO+ (Canada), with 1 kiloton of liquid scintillator. This list can
be extended to smaller detectors (less than 1 kiloton) such as Borexino
(Italy) and Baksan (Russia).  The ASPERA roadmap~\cite{2008arXiv0804.1500S},
which projects the future of astroparticle physics in Europe, includes a megaton
neutrino detector that should be operational simultaneously with
ET. (Large-scale prototypes such as MEMPHYS, LENA and GLACIER will serve as
pathfinders.) Elsewhere in the world similar projects are under development,
such as the DUSEL LBNE detector, or have entered early design phase, such as
Hyper-Kamiokande \cite{nakamura03} in Japan.  There are also more advanced
proposals for multi-megaton detectors such as Deep-TITAND
\cite{Suzuki:2001rb,Kistler:2008us}.

A similar process is exploited for the neutrinos in the high-energy range: A
charged particle results from the interaction of the neutrino and the detector
environment. The flash of Cherenkov light generated by the muon (preferred
because it travels straight through the detector, thus leaving a distinct trace)
is detected and provides evidence of the impinging neutrinos. Given the low
expected fluxes at those energies and the small cross sections, immense
instruments (km$^3$ in size) are required to detect them in sufficient
numbers. Currently, the leading experiments are IceCube, a cubic kilometer-scale
detector under construction in the ice at the geographic South Pole, and
ANTARES, which employs about $10^{-2}$ km$^3$ of sea water at 2500~m depth in
the Mediterranean sea near Marseilles (France). Looking ahead, the KM3NeT
European network recently started a design study for a km$^3$ detector in the
Mediterranean sea, which is part of the ASPERA
roadmap~\cite{2008arXiv0804.1500S}. The high-energy neutrino detectors also have
some sensitivity to neutrinos in the low-energy range; this is the case with
IceCube and a possibility under study with ANTARES. Experiments which rely upon
radio-based detection of the highest energy neutrinos~\cite{2009NIMPA.602..279B}
through the Askaryan effect, enhancing IceCube's sensitivity, might operate
together with ET -- presenting potentially interesting possibilities for
multimessenger science at the GZK frontier~\cite{2009astro2010S..43C}.

\subsection{Low-energy neutrinos}

As discussed previously, core-collapse supernovae are potential sources
of gravitational waves.  They also have a well-established low-energy
neutrino signature, and so are the flagship candidate for coincident
GW-neutrino detection. The time delay between the neutrino pulse and
the gravitational-wave emission is very small, usually assumed to be
sub-second. The almost simultaneous gravitational wave and low-energy
neutrino signal is followed by the optical signal that starts to rise
after a several-hour delay. Therefore, both gravitational-wave and
neutrino signals can be used as an early warning for electromagnetic
observers.

The core collapse of a massive star produces an O(0.1 s) long pulse of
low-energy neutrinos and antineutrinos, amounting to a total energy release
of up to $\sim$10$^{53}$~ergs.  The neutrinos produced are of all flavours,
and have energies in the few tens of MeV range. SN1987A, a supernova that
occurred at 51.4~kpc from Earth, provides us with an
observational example from which we can extrapolate our expected neutrino signal:
\begin{equation}\label{eq:scaling}
%    N ~ \simeq ~ 4.6 ~ \left(\frac{51.4 \mathrm{kpc}}{D}\right)^{2} ~ \left(\frac{M}{1 \mathrm{kton}}\right)
    N  ~ \simeq ~ 4.6 \, \left(\frac{50\,\mathrm{kpc}}{D}\right)^{2} \left(\frac{M}{1\,\mathrm{kt}}\right),
\end{equation}
where $N$ is the approximate number of neutrino interactions expected,
$D$ is the distance to the supernova and $M$ is the fiducial mass of
the low-energy neutrino detector.

A total of 24 neutrinos from SN1987A were observed by the Kamiokande-II (Japan)
\cite{hirata87:_obser_sn198}, IMB (US) \cite{bionta87:_obser_sn198_large_magel_cloud}, 
and Baksan (USSR) \cite{alekseev88:_detec_of_the_neutr_signal} detectors. Present and
future neutrino detectors are much more sensitive. For example, Super-Kamiokande
(SK) would observe $\sim$230 neutrino interactions from the same event and a
megaton detector, several thousand.

Currently, the distance reach of the global network of neutrino detectors
(including SK, LVD and SNO+) is of order O(100~kpc). Equation (\ref{eq:scaling})
indicates that for a 50~kt detector like Super-Kamiokande the $N \gtrsim 1$ 
reach is about the distance to M31, $\sim$770~kpc. Super-Kamiokande requires at
least 2 neutrino interactions in coincidence to reject background; consequently
the chance of it detecting a SN in M31 on its own is estimated at around $\sim$8\%.
This reach will gain a factor of $\sim$5 to O(1 Mpc) with the advent of megaton
class detectors~\cite{debellefon-2006,2003nipb.conf..307N,jung-2000,autiero-2007-0711}
by the ET era, thus allowing the observation of neutrinos from M31 and M33
\cite{Kistler:2008us,PhysRevLett.95.171101}. A future neutrino detector fiducial mass
increase reaching $\sim$5 Megatons, the size of the proposed Deep-TITAND, would permit
observations at a significantly larger distance, where the $N \gtrsim$1 reach is
about $\sim$8~Mpc. At this distance the
supernova rate estimate\footnote{Since 1999, 20 core-collapse supernovae have been observed
within 10 Mpc.} can be as large as 1 yr$^{-1}$ (see \cite{Kistler:2008us} and Fig.~\ref{fig:SN}).  It is thus
reasonable to expect at least one such event during the lifetime of ET.

Significant uncertainties exist in the modelling of supernovae (see also
Section\,\ref{subsection:Optical_Supernovae}) and their gravitational wave
signature. The sophisticated simulations of the core collapse of massive stars
do not robustly lead to the supernova explosion \cite{Burrows:2005dv}.  This
indicates that important physics may be lacking from the models; moreover it
seems that the electromagnetic observations alone cannot reveal the missing part
of the puzzle. Gravitational waves and neutrinos carry important information
from the innermost part of the exploding star, on the physical parameters that
govern the dynamics (e.g., degree of non-axisymmetry, rotation, magnetic field).
Those parameters might be extracted from multimessenger observations.  The exact
impact of those observations is directly connected to the availability of a
comprehensive model and a complete simulation of the core collapse process. A
pathfinder effort\cite{Leonor:2010} is currently conducted based on SNEWS
(SuperNova Early Warning System) alerts \cite{2004NJPh....6..114A} for a nearby
supernova. Due to the relatively low event rate and limited reach of both
low-energy neutrino and GW detectors for observing core-collapse supernovae, the
boost in sensitivity and detection confidence that comes from a multimessenger
approach will be a significant positive development of the ET era.

\subsection{High-energy neutrinos}

The processes leading to the production of high-energy neutrinos are tightly
connected with those of high-energy photons.  Indeed, three of the sources
discussed in Sec.~\ref{section:High-energy photons} -- gamma-ray bursts, soft
gamma-ray repeaters, and microquasar flares -- have also been identified as potential sources
of high-energy neutrinos \cite{2005ApJ...633.1013I,Pradier:2008uj}. In this
section we will focus our attention on GRBs, complementing the discussion
of Sec.~\ref{section:gamma_ray_bursts} by considering their neutrino emission.

GRBs are thought to be strong neutrino emitters.  The gamma-ray and afterglow
radiation are likely to be emitted from relativistic electrons accelerated in
shock waves; the same shocks should also accelerate protons. The protons then
collide with gamma-ray photons to produce charged pions that decay into
neutrinos
(10$^{5}$-10$^{10}$~GeV)~\cite{1997PhRvL..78.2292W,1998PhRvL..80.3690V,Waxman00}. This
scenario was first investigated for the internal shock
model~\cite{1997PhRvL..78.2292W,Piran2005}, and it has been pointed out that a km-scale
neutrino detector would observe at least several tens of events per year
\cite{Waxman00} correlated with GRBs. Many subsequent studies
(e.g.,~\cite{1998PhRvD..58l3005R,2000PhRvD..62i3015A,2000ApJ...541..707W}
% can't get those refs
%  2001PhRvL..87q1102M,2003PhRvL..90t1103G,2003PhRvL..90x1103R,2003PhRvD..68h3001R,
%  2003PhRvL..91g1102D,2006PhRvL..97e1101M,2006ApJ...651L...5M}
) have resulted in a range of different models and predictions that will be
tested against the observational data during the coming years \footnote{Some of
  the models have been already excluding thanks to the observations made by
  AMANDA \cite{2008ApJ...674..357A}.}.

We consider now in turn the two main classes of GRBs, 
long and short, starting with the former. As discussed in
Sec.~\ref{section:High-energy photons}, observed long GRBs are located at
cosmological distances. In the collapsar model (e.g.,~\cite{collapsar,Wo:93}),
they are associated with the collapse of a massive, rapidly rotating star to
a black hole.  The instabilities that develop during the collapse
may generate gravitational waves \cite{2003ApJ...589..861K,Kotake06,Ott2008}.
Several models, along with estimates of their associated GW strength, were
described in Sec. \ref{section:gamma_ray_bursts}.  Although the GW
signature is model-dependent, we can expect a distance reach to order of 1 Gpc
with ET under some scenarios.

The extensively studied GRB 080319B provides us with a case study
on searching for neutrino emission from long GRBs.  GRB 080319B,
which had a duration of 66 s, occurred at $z=0.937$ (i.e., at a luminosity
distance $\sim$6~Gpc~\cite{2009ApJ...691..723B,1999astro.ph..5116H}).
It was exceptionally bright, with $E_{\gamma}^{\mathrm{iso}}=10^{54}$ erg,
compared to $10^{53}$ erg for typical GRBs.  It is expected
\cite{Abbasi:2009kq} that such an event would be associated with
O(1) neutrino interaction in IceCube in its complete km$^3$
configuration.  Scaling down to a typical long GRB at 1 Gpc,
we expect a few neutrino events since the decrease in the
distance compensates for the lower energy.

%GRB080319B is a long (66 s duration) GRB observed recently has been studied
%extensively.  GRB080319B occurred
%at $z=0.94$ (i.e., $\sim$ 6.4 Gpc). It provides us with a case study from which
%we can infer what to expect from a typical GRB. However, note that GRB080319B
%should not be considered as typical as it is exceptionally bright (with
%$E_{\gamma}^{\mathrm{iso}}=10^{54}$ erg to be compared to $10^{53}$ erg
%typically). It is expected \cite{Abbasi:2009kq} that such event would be
%associated with O(1) neutrino interaction in Icecube (in a complete km$^3$
%configuration). When scaling down to a typical long GRB at 1 Gpc,
%we should get roughly the same number of events since
%the factor in the distance compensates the larger energy.

High-energy neutrinos could also be emitted from short-hard GRBs
\cite{Nakar06,Bloom07,Lee07,2009PhRvD..79d4024E}. We recall that the short GRBs
are thought to be associated with neutron star---neutron star or neutron
star---black hole mergers. The GW signature emitted by such binaries is
detectable to typical distances of $z\sim 2-4$ by ET. There is no widely
accepted prediction concerning the flux of high-energy neutrinos. 

However, an estimate of the neutrino flux associated with internal shocks
(i.e., interactions of $10^{16}$ eV cosmic rays with the prompt gamma-rays) can
be obtained \cite{Nakar06} assuming that the efficiency of this process is
similar to that of long GRBs. Following the Waxman and Bahcall model
\cite{1997PhRvL..78.2292W} (which estimates that in optimal conditions about 10
\% of the burst energy is emitted in the form of $10^{14}$ eV neutrinos) and
considering standard values for the jet charateristics \cite{Nakar06} (with
burst energy $\sim 10^{50}$ erg, typical duration $1$ s and Lorentz factor $\sim
30$), a source at a distance of 200 Mpc would produce $\sim$ 4 km$^{-2}$ upward
muons.

In conclusion, we should expect to see coincident detections of high-energy
neutrinos and GWs associated with both long and short GRBs during ET's lifetime if a
km$^{3}$ neutrino detector operates concomitantly. Such observations would
improve our understanding of the details of astrophysical processes connecting
the gravitational collapse/merger of compact objects to black-hole formation as
well as to the formation of fireballs.

This conclusion also applies to the low-luminosity GRBs, a subclass of
long GRBs with gamma-ray luminosities a few orders of magnitude smaller than
those of conventional GRBs (already presented in Sec. \ref{section:gamma_ray_bursts}).
Low-luminosity GRBs are associated with the lower end of the
continuum of sources of long GRBs.  Significant emission of high-energy neutrinos
is expected for those sources
~\cite{2006ApJ...651L...5M,2007APh....27..386G,2007PhRvD..76h3009W}.

%As these events are associated with an exceptionally energetic, possibly
%rapidly-rotating and jet-driven population of supernovae
%(e.g.,~\cite{Burrows2007,Dessart2008}), the GW signals might be accordingly
%stronger, making this source population an interesting target of study.

%\paragraph{Failed Gamma-Ray Bursts -- Jet-driven Supernovae}

In addition to the high- and low-luminosity GRB populations, neutrino
detectors will provide access to another subclass of GRBs that are
largely inaccessible to EM observations.
``Failed'' gamma-ray bursts are thought to be associated with baryon-rich jets.
In the presence of baryons (heavy particles), the jet cannot 
reach ultra-relativistic velocities. In such cases, the relativistic 
space-time dilation which decreases the density of
$\gamma$ photons in the jet does not occur. Over a critical density, photons
are annihilated through gamma-gamma interaction. The jet is optically thick:
no gamma-ray photons can escape.
For this reason, this class of bursts might be more challenging to observe through
conventional astronomical telescopes. GWs and high energy neutrinos can be significant sources
of information to reveal the properties of these elusive objects.

The observation of late-time radio emission by some type Ic supernovae suggests the
existence of such mildly relativistic ($\Gamma \lesssim 5$) jets~\cite{soder,Soderberg:2009ps,granot,mazzali}.
The fraction of all core-collapse supernovae with jets could perhaps
be as large as $\sim$1--10\%, and the occurrence rate of failed GRBs
is estimated to be $\sim$1--10 yr$^{-1}$ within 30 Mpc.
Ando and Beacom~\cite{2005PhRvL..95f1103A} find that, for an object
at a distance of 30 Mpc  producing a jet with kinetic energy of $3\times 10^{51}$ erg
and Lorentz factor of 3, one would expect $\sim$3 neutrino events for
a km$^3$ detector. To our knowledge, there is no specific study providing
an estimate of the GW emission by such an object. However,
since the progenitor is similar in nature to that of long GRBs, 
we may expect them to be sources of GWs accessible to ET, hence
providing another candidate for multimessenger studies with high-energy neutrinos.

\section{Discussion} \label{section:Discussion}

Interesting results from multimessenger searches using interferometric GW 
data have already been published \cite{GRB030329,SGRpaper,070201,abbott:211102,S2S3S4,Collaboration:2009zd,4051k,2009arXiv0908.3824L}.

The LIGO and VIRGO detectors have made specific scientific statements for 
some nearby events; for example, constraining the source type or location  
of GRB070201 -- a short-hard GRB event observed to come from a direction 
overlapping M31. These detectors are currently taking data near or beyond their initial 
design sensitivities, with planned upgrades to ``advanced'' configurations 
in the next few years~\cite{2009RPPh...72g6901A,2009astro2010T..51W}. 
One result is growing interest from the external astrophysics community 
in multimessenger observations with gravitational waves 
\cite{2009astro2010S.235P,2009arXiv0902.1527B,2009astro2010S.165K,2006ApJ...637...27K}. 
The anticipated
further improvement in their sensitivity gives us confidence that associations between GWs and their electromagnetic 
counterparts will be confirmed during the lifetime of the advanced gravitational detectors. Undoubtedly these 
exciting -- albeit somewhat opportunistic -- discoveries will have to be followed up via systematic studies with better 
statistics. This is the clear and fundamental promise of the ET era: the Einstein Telescope has the potential to turn 
rare discoveries into routine observations, enabling precise measurements we could not carry out before and facilitating 
population studies of GW sources. In order to fully exploit the scientific benefits offered by multimessenger 
observations, however, it will be essential that ET operates alongside partner satellites, observatories and telescopes 
across the entire electromagnetic spectrum -- from radio waves to gamma-rays -- and neutrino detectors sensitive from 
low to ultra-high energies.

To conclude, we briefly summarise some of the key astrophysical questions 
which multimessenger observations in the ET era may address.

\begin{itemize}
\item Some models predict that ET's reach for long GRBs could be as large as $\sim$O(1~Gpc). The gamma-ray 
emission and afterglow of long GRBs only provide an indirect indication of the nature of the GRB engine. By 
contrast gravitational waves produced in the immediate vicinity of the central engine, and detected by ET, 
will for the first time offer a direct probe of its physics. Furthermore ET should be able to decisively test 
the validity of predictions that a substantial population of under-luminous long GRBs are found within 
$\sim$O(100~Mpc). ET may also provide insight on the nature of XRFs and their relationship to long GRBs.

\item ET should be able to detect of order $O(10^5)$ gravitational wave sirens -- coalescing binary neutron
star systems -- in the redshift range $2 - 4$, providing precise estimates of their masses and luminosity
distances. Multimessenger observations of a subset of these systems, identified as short duration GRBs,
should allow measurement of their redshifts. These data can provide a completely independent way to calibrate
the cosmological distance scale and constrain cosmological models  -- in a manner complementary to other,
electromagnetic, cosmological probes.

\item During ET's lifetime we should expect to see coincident detections of high-energy neutrinos and GWs
associated with both long and short GRBs, if a km$^{3}$ neutrino detector operates concurrently. Such 
observations would improve our understanding of the details of astrophysical processes connecting the 
gravitational collapse/merger of compact objects to black-hole formation as well as to the formation of 
fireballs. ``Failed'' gamma-ray bursts are difficult to observe with conventional astronomical 
telescopes. GWs and high energy neutrinos can be significant sources of information to reveal the properties 
of these elusive objects.

\item At frequencies of 1-3~kHz ET will be sensitive to GW emissions from giant SGR flares, at a level as 
low as 0.01\% to 1\% of the energy content in their EM emission.  In the region of $20 - 100$ Hz, ET will be able
to probe SGR gravitational wave emissions as low as $10^{-7}$ of the total energy budget.

\item ET may be able to confirm or refute the existence of the putative high redshift population of IMBHs 
predicted in some `seed' black hole formation models. These observations may, therefore, help to discriminate 
between the various scenarios proposed to explain the origin of Ultra-Luminous X-ray binaries.

\item More generally ET will provide precise timing and acceptable sky localisation to prompt searches for
the associated optical and radio signatures of GW sources, enabling early observations that are traditionally 
opportunistic and regularly missed today.

\item Multi-megaton low-energy neutrino detectors such as Deep-TITAND
would permit observation of low energy neutrinos from a supernova at $\sim$8~Mpc. 
Within this distance the supernova rate could be as large as 1 yr$^{-1}$. 
It is thus reasonable to expect at least one such event 
during the lifetime of ET.  Multimessenger observations could reveal crucial 
information from the innermost part of the exploding star, such as the role of 
rotation and magnetic fields.
\end{itemize}

While the anticipated benefits of multimessenger observations with ET-class sensitivity are greatly encouraging in
themselves, one should not forget about the unexpected. Humanity has never aimed a new kind of telescope at the sky 
without it revealing surprises, in the form of completely new types of astrophysical objects not imagined before.
Multimessenger observations were often the key to disentangling the nature of these surprise discoveries, and we
can confidently expect that ET will play a similar role in exploring the new astrophysical frontiers of the next
decade.

\section{Acknowledgments}

%no need: ECM was supported by ???.
The research leading to these results has received funding from the
European Community's Seventh Framework Programme (FP7/2007-2013) under grant
agreement n 211743.
MH was supported in part by STFC grant PP/F001118/1.
PJS was supported in part by STFC grant PP/F001096/1.
SM was supported by the National Science Foundation under grants
PHY-0457528/0757982, PHY-0555628, and by Columbia University in
the City of New York.
We are indebted to many of our colleagues for fruitful discussions,
in particular Isabel Leonor, Ben Owen, Francois Lebrun,
Bruny Baret and Alessandra Tonazzo for their valuable suggestions.
This paper has been assigned LIGO Document Number {LIGO-P09}00100-v2. % please leave the funny brackets here -- they're needed to avoid mangling by arXiv

\bibliographystyle{spmpsci}
\bibliography{References}

% .............................................

\end{document}